\documentstyle[amssymb,preprint, aps]{revtex}

\begin{document}
\draft
\title{Phase-dependent Spontaneous Spin Polarization and Bifurcation Delay in
Coupled Two-Component Bose-Einstein Condensates}
\author{Chaohong Lee$^{\thanks{%
Corresponding author: chlee@mpipks-dresden.mpg.de and chleecn@hotmail.com.}}$}
\address{Max Planck Institute for the Physics of Complex Systems, Noethnitzer Str.
38, D-01187 Dresden, Germany, and \\
Wuhan Institute of Physics and Mathematics, The Chinese Academy of Sciences,
Wuhan 430071, P. R. China}
\author{Wenhua Hai}
\address{Department of Physics, Hunan Normal University, Changsha 410081,P. R. China}
\author{Lei Shi, and Kelin Gao}
\address{Wuhan Institute of Physics and Mathematics, The Chinese Academy of Sciences,
Wuhan 430071, P. R. China}
\date{\today}
\maketitle

\begin{abstract}
The spontaneous spin polarization and bifurcation delay in two-component
Bose-Einstein condensates coupled with laser or/and radio-frequency pulses
are investigated. We find that the bifurcation and the spontaneous spin
polarization are determined by both physical parameters and relative phase
between two condensates. Through bifurcations, the system enters into the
spontaneous spin polarization regime from the Rabi regime. We also find that
bifurcation delay appears when the parameter is swept through a static
bifurcation point. This bifurcation delay is responsible for metastability
leading to hysteresis.
\end{abstract}



\section{Introduction}

Electronic and nuclear-spin polarization in an atomic vapor with optical
pumping have been investigated extensively\cite{Fortson}. Under conditions
in which electronic spin exchange takes place faster than spin relaxation,
spontaneous spin polarization appears. This interesting phenomenon is very
similar to ferromagnetism and has been observed in wide ranges of atomic
intensity, pump laser frequency and intensity. The appearance of spontaneous
spin polarization means that the atomic vapor has two stable states with
large spin polarization. The experimental realization of it has been applied
to the field of optical bistability\cite{Gibbs}. The atomic spin
polarization exhibits striking hysteresis in switching between the bistable
states\cite{Fortson}. This is analogs to those ferromagnetic system displays
magnetic hysteresis\cite{Rao}.

With more and more deeply exploring the mechanism of the spontaneous spin
polarization phenomena in thermal atomic gases, it raises a question of
whether the spontaneous spin polarization in ultracold atomic gases is as
same as the one in thermal gases. Recently, the experimental realization of
multi-component Bose-Einstein condensates (BECs)\cite{Stenger,Hall} in
different hyperfine levels causes our interest in considering the similar
behavior of the ultracold atomic gases. There are many distinguishable
differences between thermal atomic gases and cold ones. The first one is
that the collision among thermal atomic gases is noncoherent. However, when
the temperature is close to the critical temperature realizing Bose-Einstein
condensate (BEC) ($T\thicksim T_{BEC}$) or below it, the collision among
ultracold atomic gases is coherent due to the path between such collision is
smaller than the phase coherence length\cite{Burnett,Anglin}. Can this
coherent property play a important role in the polarization process of the
ultracold atoms? Another distinct difference is that the interaction
strength of cold atoms can be controlled easily\cite{Burnett,Inouye}. But
the interaction strength in the thermal case is very difficult to control.

In this article, we shall show the coherence among ultracold atoms gives
rise to the phase-dependent spontaneous spin polarization and bifurcation
which depend on both the physical parameters and the relative phase. In the
next section, using the variational approach, we gain the model from the
mean-field description of laser coupled Bose-Einstein condensates in
different hyperfine levels of the same kind of atoms. Then, based upon the
obtained model, we analyze the parameter-dependence and the phase-dependence
of the spontaneous spin polarization from the points of bifurcation. We
simultaneously find that bifurcation delay, which is relative to
bistability/metastability, appears when related parameters slowly sweep
through static bifurcation points. Lastly, we take a brief look at
experimental possibilities. In the third section, we summarize and discuss
the obtained results.

\section{Phase-dependent Spontaneous Spin Polarization and Bifurcation Delay}

We consider that the same kind of bosonic atoms, which are trapped in a
single-well potential, are condensed in two different hyperfine levels $|1>$
and $|2>$. Raman transitions or two-photon transitions between two hyperfine
states are induced by the laser fields with the effective Rabi-frequency $%
\Omega $ and a finite detuning $\delta $. The internal Josephson effects\cite
{Stenger,Hall,Kohler,Williams,Lee,Kuang}, coherent coupling effects\cite
{coupling}, vortices\cite{Zoller} and spin waves\cite{spinwave} in such
systems have stimulated great interest of many theoretists and
experimentalists. In the rotating frame, neglecting damping and
finite-temperature effects, this coupled two-component BECs system can be
described by a pair of coupled GPEs\cite{Williams,coupling,Zoller} 
\begin{equation}
\begin{array}{l}
i\hbar \frac{\partial \Psi _{2}(\stackrel{\rightharpoonup }{r},t)}{\partial t%
}=(H_{2}^{0}+H_{2}^{MF}-\frac{\delta }{2})\Psi _{2}(\stackrel{%
\rightharpoonup }{r},t)+\frac{\Omega }{2}\Psi _{1}(\stackrel{\rightharpoonup 
}{r},t), \\ 
i\hbar \frac{\partial \Psi _{1}(\stackrel{\rightharpoonup }{r},t)}{\partial t%
}=(H_{1}^{0}+H_{1}^{MF}+\frac{\delta }{2})\Psi _{1}(\stackrel{%
\rightharpoonup }{r},t)+\frac{\Omega }{2}\Psi _{2}(\stackrel{\rightharpoonup 
}{r},t).
\end{array}
\end{equation}
Here, the free evolution Hamiltonians $H_{i}^{0}=-\frac{\hbar
^{2}\triangledown ^{2}}{2m}+V_{i}(\stackrel{\rightharpoonup }{r})$ $(i=1,2)$%
, the mean-field interaction Hamiltonians $H_{i}^{MF}=\frac{4\pi \hbar ^{2}}{%
m}(a_{ii}|\Psi _{i}(\stackrel{\rightharpoonup }{r},t)|^{2}+a_{ij}|\Psi _{j}(%
\stackrel{\rightharpoonup }{r},t)|^{2})$ $(i,j=1,2$, $i\neq j)$  and $a_{ij}$
is the scattering length between states $i$ and $j$ which satisfies $%
a_{ij}=a_{ji}$. Under the condition of weak coupling, i.e., $\Omega /\sqrt{%
\omega _{x}^{2}+\omega _{y}^{2}+\omega _{z}^{2}}\ll 1$ (in which $\Omega $
is the Rabi frequency, $\omega _{x}$, $\omega _{y}$ and $\omega _{z}$ are
the trapping frequencies), the macroscopic wavefunctions can be written in
form of the variational ansatz $\Psi _{i}(\stackrel{\rightharpoonup }{r}%
,t)=\psi _{i}(t)\Phi _{i}(\stackrel{\rightharpoonup }{r})$ $(i=1,2)$ with
amplitudes $\psi _{i}(t)=\sqrt{N_{i}(t)}e^{i\alpha _{i}(t)}$ and spatial
distributions $\Phi _{i}(\stackrel{\rightharpoonup }{r})$. In this ansatz,
the complex function $\psi _{i}(t)$ contains all time-dependence in the
macroscopic wave-function $\Psi _{i}(\stackrel{\rightharpoonup }{r},t)$ and
does not depend on the spatial coordinates. The symbols $N_{i}(t)$ and $%
\alpha _{i}(t)$ are the atomic population and phase of the $i$-th component,
respectively. Due to the coupling is very weak, the spatial distributions
vary slowly in time and are very close to the adiabatic solutions to the
time-independent uncoupled case for GP equations $(1)$, being slaved by the
populations\cite{Williams}. Thus, the amplitudes obey the nonlinear two-mode
dynamical equations 
\begin{equation}
\begin{array}{l}
i\hbar \frac{d}{dt}\psi _{_{2}}(t)=(E_{2}^{0}-\frac{\delta }{2}+U_{22}|\psi
_{_{2}}(t)|^{2}+U_{21}|\psi _{_{1}}(t)|^{2})\psi _{_{2}}(t)+\frac{K}{2}\psi
_{_{1}}(t), \\ 
i\hbar \frac{d}{dt}\psi _{_{1}}(t)=(E_{1}^{0}+\frac{\delta }{2}+U_{11}|\psi
_{_{1}}(t)|^{2}+U_{12}|\psi _{_{2}}(t)|^{2})\psi _{_{1}}(t)+\frac{K}{2}\psi
_{_{2}}(t).
\end{array}
\end{equation}
The parameters $E_{i}^{0}=\int \Phi _{i}(\stackrel{\rightharpoonup }{r}%
)H_{i}^{0}\Phi _{i}(\stackrel{\rightharpoonup }{r})d\stackrel{%
\rightharpoonup }{r}$, $U_{ij}=\frac{4\pi \hbar ^{2}a_{ij}}{m}\int |\Phi
_{i}(\stackrel{\rightharpoonup }{r})|^{2}|\Phi _{j}(\stackrel{%
\rightharpoonup }{r})|^{2}d\stackrel{\rightharpoonup }{r}=U_{ji}$ and $%
K=\Omega \int \Phi _{1}(\stackrel{\rightharpoonup }{r})\Phi _{2}(\stackrel{%
\rightharpoonup }{r})d\stackrel{\rightharpoonup }{r}$ $(i,j=1,2)$. The terms
in $K$ describe the internal tunnelling between two BEC states. Whereas the
terms in $U_{ij}$, which depend on the numbers of atoms in each BEC state,
describe the mean-field interaction between atoms. When $U_{21}$ and $\delta 
$ equals zero, these coupled equations can also describe the BECs in a
double-well potential and a nonlinear dimer\cite{doublewell}. Introducing a
Bloch spin vector with following components 
\begin{equation}
u=\frac{\psi _{_{2}}^{*}\psi _{_{1}}+\psi _{_{2}}\psi _{_{1}}^{*}}{\psi
_{_{1}}^{*}\psi _{_{1}}+\psi _{_{2}}^{*}\psi _{_{2}}},\text{ }v=-i\frac{\psi
_{_{2}}\psi _{_{1}}^{*}-\psi _{_{2}}^{*}\psi _{_{1}}}{\psi _{_{1}}^{*}\psi
_{_{1}}+\psi _{_{2}}^{*}\psi _{_{2}}},\text{ }w=\frac{\psi _{_{2}}^{*}\psi
_{_{2}}-\psi _{_{1}}^{*}\psi _{_{1}}}{\psi _{_{1}}^{*}\psi _{_{1}}+\psi
_{_{2}}^{*}\psi _{_{2}}}.
\end{equation}
Obviously, $u^{2}+v^{2}+w^{2}=1$. When the total atomic numbers $N_{T}$ $=$ $%
N_{1}+N_{2}$ $=$ $\psi _{_{1}}^{*}\psi _{_{1}}+\psi _{_{2}}^{*}\psi _{_{2}}$
is conserved, setting the Planck constant $\hbar =1$, the Bloch spin
components satisfy 
\begin{equation}
\frac{d}{dt}\left( 
\begin{array}{l}
u \\ 
v \\ 
w
\end{array}
\right) =\left( 
\begin{array}{ccc}
0 & \gamma +Gw & 0 \\ 
-(\gamma +Gw) & 0 & K \\ 
0 & -K & 0
\end{array}
\right) \left( 
\begin{array}{l}
u \\ 
v \\ 
w
\end{array}
\right) .
\end{equation}
In above Bloch equation, the parameters satisfy $\gamma
=E_{2}^{0}-E_{1}^{0}+N_{T}(U_{22}-U_{11})/2-\delta $ and $%
G=N_{T}(U_{22}+U_{11}-2U_{12})/2$. Comparing the above equation with the one
for the linear case $(U_{ij}=0)$ of equation $(2)$, one can find that the
mean-field interaction induces a shift $Gw$ in the transition frequency and
this shift is apparently proportional to the relative population $w$.

Taking $|1>$ as spin-up state and $|2>$ as spin-down state, the above
two-component BECs system can be regarded as an ensemble of quantum spin-$%
1/2 $ particles. Thus, the longitudinal component $w$ of the pseudospin
describes the relative population, and the transverse components $u$ and $v$
characterize the coherence. In this language, the effective Rabi frequency
causes an effective transverse magnetic field $K$ along axis-$u$, the
effective detuning induces an effective longitudinal magnetic field $\gamma $%
, and the mean-field interaction brings an effective longitudinal magnetic
field $Gw$ which depends on the longitudinal spin component. If one
introduce a spin $\stackrel{\rightharpoonup }{S}=(u,v,w)$ to describe the
coupled two-component BECs, the corresponding motion equation depicts the
evolution can be written as 
\begin{equation}
\frac{d\stackrel{\rightharpoonup }{S}}{dt}=-\stackrel{\rightharpoonup }{S}%
\times \stackrel{\rightharpoonup }{B}_{eff},
\end{equation}
where, the effective magnetic field $\stackrel{\rightharpoonup }{B}%
_{eff}=(-K,0,-\gamma -Gw)$.

From definition of the Bloch spin components, we know that the above system
can be described with only two independent variables. If we use the
longitudinal spin component $w$ and the relative phase $\phi =\alpha
_{2}-\alpha _{1}$ as independent variables, rescaling the time $Kt$ to $t$,
the motion equations 
\begin{equation}
\begin{array}{l}
dw/dt=-\sqrt{1-w^{2}}\sin \phi , \\ 
d\phi /dt=-\gamma /K-(G/K)w+w\cos \phi /\sqrt{1-w^{2}}.
\end{array}
\end{equation}
are equivalent to the Bloch's equation. The above equations are consistent
with those derived from the secondary quantized model\cite{Kohler}. The form
of the motion equation (6) is similar to the one for the condensates in a
double-well potential coupled with tunnelling\cite{doublewell}. However, due
to the difference between the original model, the physical means of the
parameters are very different. For the condensates in a double-well
potential, two condensates are well spatially separated, thus the mean-field
interaction between two condensates can be ignored; but for the case of
coupled two hyperfine-level condensates, the inter-condensate mean-field
interaction play a very important role due to their significant overlap. One
distinct result induced by this difference is the sign of the parameter $G$,
for the case of condensates in a double-well potential, the sign is just
determined by the sign of the scattering length; but for the other case, the
sign is determined by the balance between the internal condensate and
inter-condensate mean-field interactions.

\subsection{Phase-dependent Spontaneous Spin Polarization}

Before analyze the spontaneous spin polarization in coupled two-component
BECs, we give a brief review of the spontaneous spin polarization in thermal
gases and the spontaneous magnetization which is very similar to the
spontaneous spin polarization\cite{Fortson,Rao}.

Usually, for a laser pumped thermal gas, if the spin exchange takes place
faster than the spin relaxation and the coupling laser only excite some
certain hyperfine levels of the ground state, spontaneous spin polarization
occurs when the laser intensity is large enough. This novel bistable
phenomenon involves three basic processes: laser pumping, spin exchange and
spin relaxation. The imbalance of transition probabilities among different
hyperfine states induced by the pumping laser will amplify the spin
polarization. The spin exchange will keep the transition imbalance and does
not destroy the spin polarization. However, the spin relaxation will
decrease the spin polarization. In the case of the rates of spin exchange
and laser pumping are larger than some certain threshold values, the thermal
gas exists two metastable non-completely polarized states. This means the
balance between spin relaxation and cooperation of laser pumping and spin
exchange.

From the viewpoints of bistability, spontaneous spin polarization is similar
to spontaneous magnetization. For a ferromagnetic system, if its temperature
is below the Curie temperature, slowly changing the magnetic field from
negative to positive, a first order phase transition occurs when the
magnetic field sweeps through the zero point. At the zero field point, two
metastable states with different spontaneous magnetization directions
appear. The appearing state depends on the initial magnetization. Increasing
the temperature above the Curie temperature, the spontaneous magnetization
disappears, this corresponds to occurrence of a second order phase
transition.

Under some certain conditions, both spontaneous spin polarization in a
thermal gas and spontaneous magnetization in a ferromagnetic system appear
novel bistability, and they just only depend on the related parameters. In
the following, we find a novel character of the spontaneous spin
polarization in laser coupled two-component BECs: phase dependence. That is,
the spontaneous spin polarization in coupled two-component BECs depends on
not only the related parameters but also the relative phase between two
condensates.

Now, let us analyze the spontaneous spin polarization in coupled
two-component BECs. It is well known, for a bounded dynamical system, the
stable behavior usually oscillate around some stationary states. This
indicates, the nonzero time-averaged value of a physical variable in a
stable evolution requires that the system possesses at least a stationary
state with nonzero value for this variable. Thus if there coexist multiple
stationary states with nonzero spin polarization $w$ appearing in the
coupled two-component BECs, the spontaneous spin polarization will appear.
This means, one can explore the behavior of spontaneous spin polarization by
analyzing the corresponding stationary states. The stationary states can be
obtained from the stable fixed points of the system, which correspond to
those solutions satisfying $dw/dt=0$ and $d\phi /dt=0$. In the region $%
[0,2\pi )$ of the relative phase, we find two different modes of stationary
states existing in the system: one is the equal-phase mode with zero
relative phase $(\phi =0)$, the other one is the anti-phase mode with $\pi $
relative phase $(\phi =\pi )$. Small oscillations around those stationary
states with nonzero spin polarization are special kinds of macroscopic
quantum self-trapping (MQST) states, which have also been found in the
condensates trapped in a double-well potential\cite{doublewell}. For the
MQST sates, they only require the states oscillate around $dw/dt=0$, thus
running-phase MQST states can appear. However, for the small oscillations
around stationary states, their centers are stationary states which satisfy
both $dw/dt=0$ and $d\phi /dt=0$, thus running-phase MQST states never
appear. From the points of stability, running-phase MQST states are
unstable, but all small oscillations around stationary states are all stable.

The number of fixed points and stationary states depend on the ratios $%
\gamma /K$, $G/K$ and the relative phase. For the equal-phase mode, only a
fixed point exists when $G/K\leq 1$ and this fixed point is stable. When $%
G/K>1$, there are two stable fixed points and an unstable one for $%
(G/K)^{2/3}-(\gamma /K)^{2/3}>1$ and only one stable fixed point for $%
(G/K)^{2/3}-(\gamma /K)^{2/3}<1$. Saddle-node bifurcations occur at the
points satisfying $(G/K)^{2/3}-(\gamma /K)^{2/3}=1$. In the left column of
Fig. 1, we show the values for the longitudinal component of the fixed
points in the equal-phase mode with different ratios $\gamma /K$ and $G/K$.
For the anti-phase mode, the parametric dependence of fixed points and
stationary states is very different. When $G/K\geq -1$, only a fixed point
appears and it is stable. When $G/K<-1$, two stable fixed points and an
unstable one exist for $(G/K)^{2/3}-(\gamma /K)^{2/3}>1$ and only one stable
fixed point emerges for $(G/K)^{2/3}-(\gamma /K)^{2/3}<1$. Saddle-node
bifurcations also occur at the points satisfying $(G/K)^{2/3}-(\gamma
/K)^{2/3}=1$. The fixed points of the anti-phase mode with different ratios $%
\gamma /K$ and $G/K$ are exhibited in the right column of Fig. 1.

In the Fig. 1, the fixed points between a pair of bifurcation points with
same ratio $G/K$ are unstable and the values for $d(\gamma /K)/dw$ at the
bifurcation points equal zero. From the previous analysis, we find
bistability exists in either the equal-phase mode or the anti-phase mode
when the parameters obey $(G/K)^{2/3}-(\gamma /K)^{2/3}>1$. The appearance
of bistability indicates the existence of spontaneous spin polarization in
this coupled two-component BECs system. When $\left| K/G\right| <1$, and $%
\gamma /K$ goes through the bifurcation points which satisfy $%
(G/K)^{2/3}-(\gamma /K)^{2/3}=1$, the spin polarization of either the
equal-phase mode or the anti-phase mode is discontinuous at the bifurcation
points. This means a first-order phase transition of the spin polarization.
It resembles that the first-order phase transition, spontaneous
magnetization, in a ferromagnetic system below the Curie temperature takes
place when vary direction of the magnetic field. The difference is that,
spontaneous magnetization only occurs at the zero field point, however,
spontaneous spin polarization occurs in the region between a pair of
bifurcation points. Increasing $\left| K/G\right| $ to $1$, the spontaneous
spin polarization vanishes, which corresponds to a second-order phase
transition of the spin polarization. This is similar to spontaneous
magnetization disappears in a ferromagnetic system when the temperature is
increased to the Curie temperature. Thus, the ratio $K/G$ corresponds to the
temperature in a ferromagnetic system and $\left| K/G\right| =1$ takes the
role of Curie temperature.

Similar to the case of thermal atoms, spontaneous spin polarization can be
induced by adjusting the coupling lasers. Additionally, due to the
collisions among ultracold atoms can be controlled easily, spontaneous spin
polarization in Bose condensed atoms can also be induced by adjusting the
collision strength through bifurcation. This seems to adjust the temperature
of a ferromagnetic system. Tuning the coupling laser with fixed intensity to
a certain detuning satisfying $\gamma =0$, the bifurcation and the
spontaneous spin polarization caused by the ultracold collisions can be
obtained. For the equal-phase mode, only one stable fixed point $w=0$ exists
if $G/K<1$ and two new stable fixed points $w_{\pm }=\pm \sqrt{1-(G/K)^{-2}}$
appear with the original one $w=0$ becomes unstable if $G/K>1$. This means a
Hopf bifurcation takes place when $G/K=1$. The system goes from the Rabi
regime $(G/K<1)$ into the spontaneous spin polarization regime $(G/K>1)$
through this Hopf bifurcation. However, for the anti-phase mode, the Hopf
bifurcation occurs at $G/K=-1$. There is only one stable fixed point $w=0$
for $G/K>-1$ and two table fixed points $w_{\pm }=\pm \sqrt{1-(G/K)^{-2}}$
with an unstable one $w=0$ for $G/K<-1$. Correspondingly, the parametric
regime sustaining spontaneous spin polarization satisfies $G/K<-1$. The Hopf
bifurcations in both equal-phase mode and anti-phase mode are shown in Fig.
2. The solid lines are stable equilibria (stationary states), the dot lines
are unstable equilibria.

\subsection{Phase-dependent Bifurcation Delay}

The bifurcations obtained by analyzing the equilibria with fixed parameters
are static bifurcations. For a real physical system, some parameters can be
accurately tuned by turning knobs of experimental apparatus. When the
parameters are swept through a static bifurcation point, an interesting
phenomenon emerges: the system starting close to the initially stable
equilibrium does not immediately react to the bifurcation. Furthermore, it
remains for some time close to the unstable equilibrium, then fast falls
into one of the newly formed stable equilibria. This has been named as
bifurcation delay which has been found in a variety of physical systems\cite
{bfdelay}. The bifurcation delay, which might lead to hysteresis, is the
response to the bistability.

Slowly varying some parameters, the coupled two-component BEC system also
exhibits the phenomenon of bifurcation delay. For the equal-phase mode,
fixed the effective detuning $\gamma =0$, slowly sweeping up the ratio $G/K$
from $R_{0}$ with sweeping rate $r$ $($i.e., $G/K=R_{0}+rt$, $1\gg r>0)$,
choosing $R_{0}<1$ and the initial state close to the stable equilibrium,
the system evolves along the unstable equilibrium for a period of time after
the ratio sweeping through the static bifurcation point $(G/K=1)$, then it
quickly goes into a small oscillations around one of two new stable
equilibria. The equilibrium, which the system evolves around lastly,
determines by the state at the static bifurcation point. The system evolves
around the up branch lastly when this state is close to the up branch;
otherwise, the system evolves around the down branch. When $R_{0}>1$, slowly
sweeping down the ratio through the static bifurcation point with initial
state close to one of two stable equilibria, the system evolves near the
stable equilibrium before it sweeps through the static bifurcation point,
then it goes into a small Rabi oscillation around the ordinary equilibrium ($%
w=0$). For the same sweeping rate, averaging the small oscillations, the
process of sweeping up and down generates a loop in the plane extended by $%
G/K$ and $w$. The area enclosed in the loop increases with the sweeping
rate. This means that the energy exchanged between the atoms and the
environments increases with the sweeping rate. The bifurcation delay in the
equal-phase mode with different sweeping rate is shown in Fig. 3. For the
anti-phase mode, a similar behavior can be observed near the static
bifurcation point $G/K=-1$.

\subsection{Experimental Possibilities}

Based upon the works of JILA\cite{Hall} and LENS\cite{LENS}, we now discuss
experimental possibilities of observing the spontaneous spin polarization
and bifurcation delay predicted in above. Using the developed experimental
technique\cite{Hall,LENS}, one can prepare two BECs in the $|F=1,m_{F}=-1>$
and $|2,1>$ hyperfine spin states of $^{87}R_{b}$ which are coupled by
introducing a two-photon pulse with the two-photon Rabi-frequency $\Omega $
and a finite detuning $\delta $. In the case of pure condensates which has
been analyzed in previous, controlling the parameters $K$ and $\gamma $ can
be realized by adjusting Rabi frequency and detuning of the coupling lasers,
respectively. Tuning the parameter $G$ can be accomplished by varying the
scattering lengths with Feshbach resonances\cite{Inouye}. The time-evolution
of longitudinal and transverse spin components can be measured with the
state-selective absorption imagining and the Ramsey interference,
respectively\cite{spinwave}.

Thus, to observe spontaneous spin polarization, one just need choose proper
fixed values for Rabi-frequency $\Omega $, detuning $\delta $ and scattering
lengths satisfying $(G/K)^{2/3}-(\gamma /K)^{2/3}>1$. To observe bifurcation
delay, one has to fix Rabi-frequency $\Omega $ and detuning $\delta $ and
slowly vary scattering lengths (or fix scattering lengths and detuning and
slowly tune Rabi-frequency) through a static bifurcation point. There are
two ways to observe these behaviors, one way is by directly observing the
stationary states, the other way is by observing small oscillations around
stationary states. Observing stationary state behavior may be not easy,
because the relaxation time of a pure condensate is much longer than that of
a thermal gas and time scale of this relaxation process, which is relevant
to real experimental systems including finite temperature thermal clouds, is
still an open problem. Fortunately, due to the averaged center of small
oscillations is very close to the surrounded stationary state, the averaged
center becomes a good understudy.

\section{Summary and Discussion}

Summary, due to the coherent ultracold collision among condensed bosonic
atoms, the bifurcation and the spontaneous spin polarization in coupled
two-component BECs rely on both relative phase and physical parameters.
These phenomena are different from those only determined by physical
parameters, we name them as phase-dependent bifurcation and phase-dependent
spontaneous spin polarization, respectively. For zero effective detuning $%
\gamma $, Hopf bifurcation and bifurcation delay can be induced by a
Feshbach resonance in either the equal-phase mode or the anti-phase mode.
The system falls into the spontaneous spin polarization regime from the Rabi
regime after a bifurcation occurs. The appearance of bifurcation delay
indicates the existence of metastability and hysteresis. Because of the
inherently quantum coherence and superposition of two condensates, this kind
of quantum metastability and hysteresis might open the door to storage
quantum data with Bose condensed atoms\cite{quanthys}.

In this article, we have focused on the novel phenomena in the case of
constant parameters. Now, we give a brief discussion about the effects of
the implicit time-dependence of the parameters. In the weak coupling case,
the spatial functions $\Phi _{i}(\stackrel{\rightharpoonup }{r})$ weakly
depend on the population difference. Thus these functions weakly rely on
time implicitly when the population difference varies with time. The
numerical results in Ref.\cite{Williams} show the overlap between two
condensates keeps nearly unchanged and the chemical potential difference is
approximately a linear function of the population difference when
populations are varied. This indicates that the parameters slowly fluctuate
around some certain constants with very small amplitudes when the population
oscillates. Thus the real population oscillation slightly depart from the
one with constant parameters.

We also note that bistability and anti-phase mode are relative to population
self-trapping states and $\pi $-states in a double-well Bose condensates or
a nonlinear dimer\cite{doublewell}. The appearance of bistability means the
existence of metastable self-trapping states. But not all self-trapping
states exhibit bistability, such as running-phase self-trapping states.
Additionally, there exist distinct difference in physical models, analysis
methods and discussed phenomena. For the physical models, two condensates in
a double-well potential are well spatially separated, thus mean-field
interaction between two condensates is negligible. However, mean-field
interaction between two hyperfine-state condensates acts an important role
due to their significant overlap. For the analysis methods, the authors of
Ref.\cite{doublewell} solve motion equations for some certain initial
conditions with numerical approach or analyze stationary states of symmetric
case. In this article, from the viewpoints of bifurcation, we exactly
analyze not only the fixed points themselves but also their stability for
arbitrary parameters. For the discussed phenomena, the works in Ref.\cite
{doublewell} are applied to population self-trapping and macroscopic quantum
tunnelling, our work firstly explores the spontaneous spin polarization and
bifurcation delay in laser pumped Bose condensed dilute atomic gases.

\begin{center}
{\large Acknowledgment}
\end{center}

The work is supported by foundations of MPI-PKS and Chinese NSFC (Grant No.
10275023 and 10274093), National Fundamental Research Program (Grant No.
2001CB309300). We thanks Dr. J. Feng of WIPM for useful comments.

\begin{center}
{\large Figure caption}
\end{center}

Fig. 1 The fixed points for the system with different ratios $\gamma /K$ and 
$G/K$. The numbers labelled on the lines are values for $G/K$.

Fig. 2 The static Hopf bifurcation and the spontaneous spin polarization.

Fig. 3 The bifurcation delay in the equal-phase mode for different values of
sweeping rate which are labelled on the lines.

\end{document}